\newcommand{\ale}{\ \raisebox{-.3ex}{$\stackrel{<}{\scriptstyle \sim}$}\ }
\newcommand{\age}{\ \raisebox{-.3ex}{$\stackrel{>}{\scriptstyle \sim}$}\ }

\documentstyle{mn}

\input psfig

\title[Evolution of layered protoplanetary discs]
	{Episodic accretion in magnetically 
	layered \\ protoplanetary discs}
	
\author[P.J. Armitage, M. Livio \& J.E. Pringle]{Philip J. Armitage$^1$
	\thanks{Present address: School of Physics and Astronomy, University 
	of St Andrews, North Haugh, St Andrews KY16 9SS}, 
        Mario Livio$^2$ and J.E. Pringle$^{2,3}$ \\
	$^1$Max-Planck-Institut f\"ur Astrophysik, 
	Karl-Schwarzschild-Str. 1,
        D-85741 Garching, Germany \\
	$^2$Space Telescope Science Institute, 3700 San Martin Drive, 
	Baltimore, MD21218, USA \\	
	$^3$Institute of Astronomy, 
	Madingley Road, Cambridge, CB3 0HA, UK}	
 	
\begin{document}

\maketitle

\begin{abstract}
We study protoplanetary disc evolution assuming 
that angular momentum transport is driven by 
gravitational instability at large radii, and magnetohydrodynamic (MHD) 
turbulence in the hot inner regions. At radii of the order of 1~AU such 
discs develop a magnetically layered structure, with accretion occurring in 
an ionized surface layer overlying quiescent gas that is too cool 
to sustain MHD turbulence. We 
show that layered discs are subject to a limit cycle instability, in 
which accretion onto the protostar occurs in $\sim 10^4 \ {\rm yr}$ bursts with 
$\dot{M} \sim 10^{-5} M_\odot {\rm yr}^{-1}$, separated by 
quiescent intervals lasting $\sim 10^5 \ {\rm yr}$ 
where $\dot{M} \approx 10^{-8} M_\odot {\rm yr}^{-1}$. 
Such bursts could lead to repeated episodes of strong mass 
outflow in Young Stellar Objects.
The transition to this episodic mode of accretion occurs at an early 
epoch ($t \ll 1 \ {\rm Myr}$), and the model therefore predicts that many young 
pre-main-sequence stars should have low rates of accretion through the inner 
disc. At ages of a few 
Myr, the discs are up to an order of magnitude more massive than 
the minimum mass solar nebula, with most of the mass locked up in the 
quiescent layer of the disc at $r \sim 1 \ {\rm AU}$. The predicted 
rate of low mass planetary migration is reduced at the outer edge 
of the layered disc, which could lead to an enhanced probability of 
giant planet formation at radii of 1 -- 3 AU.
\end{abstract}

\begin{keywords}	
	accretion, accretion discs --- MHD ---
	stars: pre-main-sequence --- stars: formation --- solar system: formation ---
	planets and satellites
\end{keywords}

\section{Introduction}

The structure and evolution of protoplanetary discs
depend upon the rate at which gas can shed its angular momentum and 
thereby flow inwards. Two widely applicable physical mechanisms
are known to lead to the required outward angular momentum 
transport. If the gas is coupled to a magnetic field, instabilities 
that inevitably arise in differentially rotating discs (Balbus \& 
Hawley 1991; Chandrasekhar 1961; Velikhov 1959) lead to turbulence and 
angular momentum transport (Stone et al. 1996; Brandenburg et al. 
1995; for a review see e.g. Hawley \& Balbus 1999). If the disc is 
massive enough, gravitational instability leads to additional transport  
(Toomre 1964; Laughlin \& Bodenheimer 1994; Nelson et al. 1998; 
Pickett et al. 2000). 

Applying these findings to the construction of protoplanetary 
disc models leads to the structure shown schematically in 
Fig.~1 (after Gammie 1996). In the inner disc, MHD turbulence 
transports angular momentum. However, at larger radii of 
$r \sim 1 \ {\rm AU}$, where the temperature is 
typically a few hundred K, magnetic field instabilities are 
suppressed by the low ionization fraction
(Matsumoto \& Tajima 1995; Gammie 1996; Gammie \& Menou 1998; 
Livio 1999; Wardle 1999; Sano \& Miyama 1999; Sano et al. 2000). This 
leads (Gammie 1996) to the formation of a {\em layered} disc 
structure, in which the gas near the disc midplane is cold,
shielded from ionizing high energy radiation, and quiescent 
(non-turbulent). Turbulence and accretion occurs only in 
a thin surface layer that is ionized by cosmic rays. Moving 
still further outwards the entire thickness of the disc 
again become viscous, either at the radius where  the surface density 
is small enough for cosmic rays to penetrate to the midplane,
or where the onset of disc self-gravity provides an 
alternative non-magnetic source of angular momentum transport.

\begin{figure}
\psfig{figure=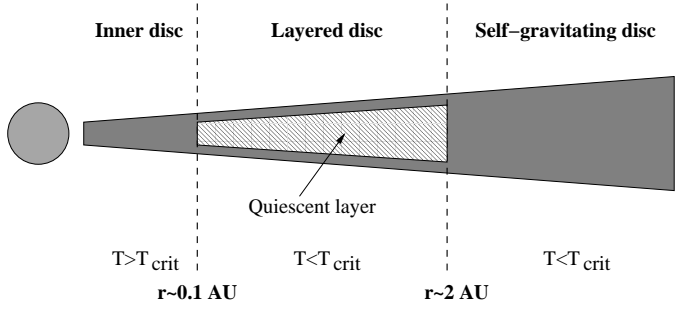,width=3.5truein,height=1.6truein}
\caption{Illustration of the radial structure of the disc in the 
	layered disc model. At small radii, 
	the central temperature exceeds $T_{\rm crit}$, the 
	temperature above which MHD turbulence provides an 
	efficient source of angular momentum transport. In the 
	intermediate, layered region, the central temperature 
	is too low for the disc to support MHD turbulence. 
	Accretion occurs through a surface layer which is 
	kept ionized by cosmic rays, while the disc near the 
	midplane is quiescent. At larger radii, the entire 
	thickness of the disc is again viscous, with angular 
	momentum transport being driven by self-gravity.}
\label{f1}
\end{figure}

The predictions of a static layered disc model for the 
accretion rate and spectral energy distribution of T Tauri 
stars were discussed by Gammie (1996), and are broadly 
consistent with observations (e.g. with the accretion 
rate for Classical T Tauri stars measured by Gullbring et al. 1998).
In this paper we consider the evolution of the layered disc,
which cannot be in a steady state (Gammie 1996, 1999; Stepinski 1999), and 
examine the implications for the outflow history of young 
stars and for the predicted disc mass. The most significant 
changes to the disc structure occur at the radii of 
greatest interest for planet formation (Reyes-Ruiz \& Stepinski 1995), 
and we discuss the implications for the migration of low 
mass planets, and for the eccentricity of massive planets interacting 
with the disc.

\section{Layered protoplanetary disc evolution}

\subsection{Equations}

Describing the evolution of the surface density $\Sigma(r,t)$ and midplane 
temperature $T_c(r,t)$ of a layered disc requires only minor modifications 
to the usual time-dependent equations for thin accretion discs. We denote 
the surface density of the `active' (viscous) disc by $\Sigma_a$. If, 
\begin{equation} 
 T_c > T_{\rm crit}
\label{eq1}
\end{equation}
or, 
\begin{equation} 
 \Sigma < 2 \ \Sigma_{\rm layer}
\label{eq2}
\end{equation}
then the disc is viscous throughout its thickness and $\Sigma_a =  \Sigma$. 
Otherwise only the surface layers are viscous and $\Sigma_a =  2 \Sigma_{\rm layer}$. 
The values of these parameters are determined by the requirement that the 
disc be sufficiently ionized to support MHD turbulence (Gammie 1996).
We adopt $T_{\rm crit} = 800 \ {\rm K}$, and $\Sigma_{\rm layer} = 10^2 \ {\rm gcm}^{-2}$. 
For a Keplerian disc, the angular velocity is $\Omega = \sqrt{GM_* / r^3}$, 
where $M_*$ is the stellar mass. 
The surface density evolution is then described by,
\begin{equation} 
 { {\partial \Sigma} \over {\partial t} } = { 3 \over r } 
 { \partial \over {\partial r} } \left[ r^{1/2} 
 { \partial \over {\partial r} } \left( 
 \nu \Sigma_a r^{1/2} \right) \right] + {\dot{\Sigma}} (r,t),
\label{eq3}
\end{equation} 
where $\nu$ is the kinematic viscosity and ${\dot{\Sigma}} (r,t)$ is the 
rate of change of the surface density due to infall onto the 
disc.

For the energy equation, we adopt a simplified form of that 
used by Cannizzo (1993),
\begin{equation} 
 { {\partial T_c} \over {\partial t} } = 
 { { 2 ( Q_+ - Q_- ) } \over {c_p \Sigma} } - 
 v_r { {\partial T_c} \over {\partial r} }.
\label{eq4}
\end{equation} 
Here $c_p$ is the disc specific heat, which for temperatures 
$T_c \sim 10^3 \ {\rm K}$ is given by $c_p \simeq 2.7 \cal{R} / \mu$, 
where $\cal{R}$ is the gas constant and $\mu=2.3$ is the mean 
molecular weight. $Q_+$ represents local heating due to 
viscous dissipation, given by,
\begin{equation} 
 Q_+ = \left( 9 \over 8 \right) \nu \Sigma \Omega^2 
\label{eq5}
\end{equation} 
if the entire disc is viscous and
\begin{equation} 
 Q_+ = \left( 9 \over 8 \right) \nu \Sigma_{\rm layer} \Omega^2 
\label{eq6}
\end{equation}
otherwise. For $Q_-$, the local cooling rate, we assume that each annulus of the disc 
radiates as a blackbody at temperature $T_e$, so that 
\begin{equation}
 Q_- = \sigma T_e^4,
\label{eq7}
\end{equation}
where $\sigma$ is the Stefan-Boltzmann constant. Finally, we 
include an advective term in the energy equation, which 
depends on the vertically averaged radial velocity,
\begin{equation} 
 v_r = - {3 \over {\Sigma r^{1/2}} }
 { \partial \over {\partial r} } \left( \nu \Sigma r^{1/2} \right),
\label{eq8}
\end{equation}
and the radial temperature gradient.

\subsection{Vertical structure}

Completing the model requires specification of both the viscosity $\nu$ and 
the vertical structure, which sets the relation between the central 
temperature $T_c$ and the surface temperature $T_e$. We adopt the 
simplest, vertically averaged approach, for which,
\begin{equation} 
 T_c^4 = { 3 \over 8 } \tau T_e^4,
\label{eq9} 
\end{equation}
where $\tau = (\Sigma / 2) \kappa$ is the optical depth for a given 
opacity $\kappa (\rho_c,T_c)$. When an annulus makes the transition 
to the layered state, we crudely account for this by replacing $(\Sigma / 2)$ 
in the expression for $\tau$ by $\Sigma_{\rm layer}$. Note that 
this means that we do not attempt 
to treat the vertical structure {\em during} the transition consistently.

\begin{table}
\begin{tabular}{clcll}
       $n$ & $\kappa_0$ & $a$ & $b$ & $T_{\rm max}$ / K \\ \hline
       
       1  & $1.0 \times 10^{-4}$ & 0 & 2.1 & 132  \\
       2  & $3.0 \times 10^{0}$ & 0 & -0.01 & 170  \\
       3  & $1.0 \times 10^{-2}$ & 0 & 1.1 & 377  \\       
       4  & $5.0 \times 10^{4}$ & 0 & -1.5 & 389  \\
       5  & $1.0 \times 10^{-1}$ & 0 & 0.7 & 579  \\
       6  & $2.0 \times 10^{15}$ & 0 & -5.2 & 681  \\       
       7  & $2.0 \times 10^{-2}$ & 0 & 0.8 & 964  \\       
       8  & $2.0 \times 10^{81}$ & 1 & -24 & 1572  \\       
       9  & $1.0 \times 10^{-8}$ & $2/3$ & 3 & 3728  \\       
       10 & $1.0 \times 10^{-36}$ & $1/3$ & 10 & 10330  \\       
       11 & $1.5 \times 10^{20}$ & 1 & $-5/2$ & 45060 \\
       12 & $0.348$ & 0 & 0 & --- \\ \hline
\label{opacity_table}
\end{tabular}
\caption{Opacity regimes in ascending order of temperature, fitted by 
         analytic functions of the form $\kappa = \kappa_0 \rho^a T^b$. 
	 We have used fits provided by Bell \& Lin (1994), as modified 
	 for low temperatures by Bell et al. (1997). The maximum 
	 temperature $T_{\rm max}$ for each regime is quoted for 
	 a typical disc density of $10^{-9} \ {\rm g cm}^{-3}$ 
	 (where specification of the density is necessary).}
\end{table}

Analytic expressions for low temperature Rosseland mean opacities are given 
by Bell et al. (1997). The behaviour of the disc depends primarily on the 
opacity near the transition temperature $T_{\rm crit}$, for which 
the fit is,
\begin{equation} 
 \kappa = 2 \times 10^{-2} T_c^{0.8} \ {\rm cm^2} \ {\rm g}^{-1}.
\label{eq10}
\end{equation} 
The full list of opacities used is quoted in Table 1.
These fits have been taken from Bell \& Lin (1994), with the modifications 
for low temperatures quoted in Bell et al. (1997).

\subsection{Viscosity}

We adopt an alpha prescription (Shakura \& Sunyaev 1973) for the 
viscosity,
\begin{equation} 
 \nu = \alpha {c_s^2 \over \Omega}, 
\label{eq11}
\end{equation}
where $c_s = \sqrt{ {\cal{R}} T_c / \mu}$ is the midplane sound speed 
and $\alpha$ a dimensionless parameter measuring the efficiency 
of angular momentum transport. Numerical simulations of MHD 
turbulence in discs suggest that $\alpha \approx 10^{-2}$ 
(Stone et al. 1996; Brandenburg et al. 1995).

\subsection{Self-gravity}

An immediate consequence of the layered disc structure is that 
the accretion rate through the surface layer is a fixed, increasing 
function of radius (Gammie 1996). Inevitably, therefore, material 
builds up in the quiescent layer over time, until the surface density 
becomes high enough that self-gravity of the disc becomes important. 
The condition for this is that the Toomre $Q$ parameter (Toomre 1964), 
given by,
\begin{equation} 
 Q = { {c_s \Omega} \over {\pi G \Sigma} },
\label{eq12} 
\end{equation}
is less than some critical value $Q_{\rm crit}$, which is of the order 
of unity. Here, we adopt $Q_{\rm crit} = 2$.
When $Q < Q_{\rm crit}$, non-axisymmetric instabilities set in 
and act to transport angular momentum outwards. In doing so, the surface 
density evolves in such a way that the stability of the disc generally 
increases, thereby self-regulating the violence of the instability.

In a local approximation, the efficiency of angular momentum 
transport due to self-gravity can be determined by the requirement 
that viscous heating balances radiative losses (Gammie 1999). This 
approach, in which the cooling time directly determines $\alpha$, 
is most appropriate for discs whose thermal balance is determined by 
viscous heating. Even in this limit, global effects may 
be important (Balbus \& Papaloizou 1999). 
A global redistribution of surface density, driven 
by self-gravity, can also occur even if the  
disc temperature is set externally (for example by irradiation). This 
is the limit that has been studied extensively using isothermal or polytropic 
simulations (e.g. Laughlin \& Bodenheimer 1994). For protoplanetary 
discs, models suggest that both viscous heating and irradiation can 
be important at different epochs and radii (Bell et al. 1997).

Given the above complications, it is evident that attempts to 
include the effects of self-gravity into one dimensional disc models 
can only be approximate. 
We adopt the form suggested by Lin \& Pringle (1987, 1990), and use 
a variant of equation (\ref{eq11}) along with an effective 
gravitational `$\alpha$',
$\nu_{\rm grav} = \alpha_{\rm grav} c_s^2 / \Omega$, with 
\begin{eqnarray} 
 \alpha_{\rm grav} & = & 0.01 \left( { Q_{\rm crit}^2 \over Q^2 } - 1 \right)
 \ \ \ {\rm if} \ Q < Q_{\rm crit} \nonumber \\
 \alpha_{\rm grav} & = & 0 \ \ \ {\rm otherwise}.
\label{eq13}
\end{eqnarray} 
Operationally, we solve equations (\ref{eq3}) and (\ref{eq4}) using an 
explicit finite difference method, and include the self-gravity as 
an additional sweep over the grid using $\nu_{\rm grav}$ and the 
full surface density $\Sigma$ in the diffusive part of equation (\ref{eq3}), 
and as an additional heating term in the energy equation.

\section{Results}

For our numerical calculations we 
adopt $M_* = M_\odot$, an inner disc radius $r_{\rm in} = 3.5 \times 
10^{11} \ {\rm cm}$ (5 $R_\odot$), and an outer disc radius 
$r_{\rm out} = 6.0 \times 10^{14} \ {\rm cm}$ (40 AU). 
The computational grid of 120 radial mesh points is uniform in a scaled radial variable 
$X \propto r^{1/2}$. A zero torque boundary condition is imposed  
at $r_{\rm in}$, while at $r_{\rm out}$ we prevent outflow by setting the radial 
velocity, $v_r$, to zero. When mass is added to the 
disc (representing infall of further material from the molecular 
cloud), we take $\dot{\Sigma}$ to be a Gaussian centered at 10 AU 
with a width of 1 AU. This is exterior to the layered section of the 
disc that we are interested in studying, so the details of how 
mass is added are unimportant here. We 
add mass assuming that it has the 
same specific angular momentum and temperature as the local disc 
material.

\subsection{Steady infall models}

To demonstrate the evolution of the layered disc model, we 
first take the infall rate 
onto the outer disc $\dot{M}_{\rm infall}$ to be a constant. We 
take $\alpha = 10^{-2}$, as suggested by numerical simulations 
of MHD disc turbulence, and allow the disc to evolve from an 
initially low mass ($10^{-2} \ M_\odot$) until either a 
steady state or a limit cycle has developed.

\begin{figure}
\psfig{figure=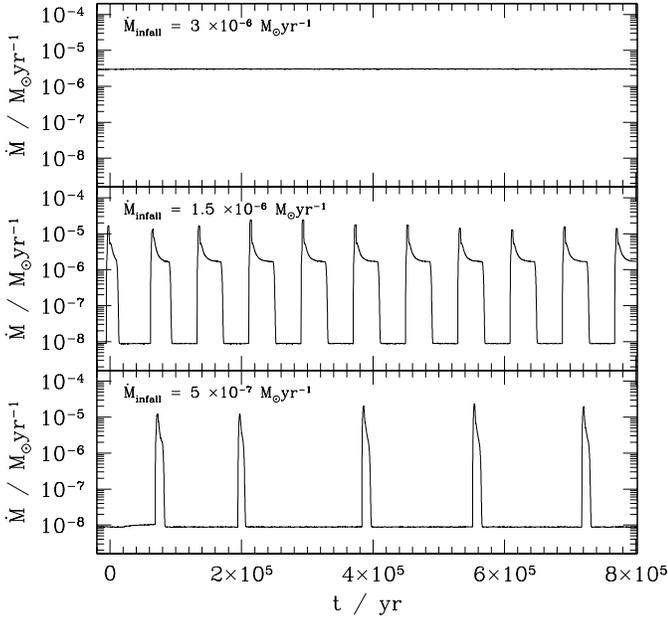,width=3.5truein,height=3.5truein}
\caption{Protostellar accretion for models with a constant rate 
         of infall onto the outer disc. From top down, the panels 
	 show models with $\dot{M}_{\rm infall} = 3 \times 10^{-6} \ M_\odot 
	 {\rm yr}^{-1}$, $\dot{M}_{\rm infall} = 1.5 \times 10^{-6} \ M_\odot 
	 {\rm yr}^{-1}$, and $\dot{M}_{\rm infall} = 5 \times 10^{-7} \ M_\odot 
	 {\rm yr}^{-1}$ respectively. All models are shown after initial 
	 transients have decayed. Steady accretion occurs for sufficiently 
	 high accretion rates through the outer disc. At lower accretion rates, 
	 the mass flow onto the star is strongly time-dependent.}
\label{f2}
\end{figure}

Figure 2 shows the accretion rate onto the star for a range of mass 
infall rates between $5 \times 10^{-7} \ M_\odot {\rm yr}^{-1}$ and 
$3 \times 10^{-6} \ M_\odot {\rm yr}^{-1}$. For the highest infall rate, 
the disc is able to transport gas steadily onto the star. The 
inner region, in which $T_c > T_{\rm crit}$, is able to be 
supplied by an outer disc in which $Q < Q_{\rm crit}$ and the 
angular momentum is transported by self-gravity. $Q$ is typically 
around 1.5 in this region, with the transition occurring at 
a radius of $\approx 3$ AU. 

For lower infall rates 
such steady accretion is not possible. After an initial phase in which 
the disc mass increases steadily, a limit cycle is obtained in which 
outbursts, with a peak accretion rate 
$\dot{M} \approx 10^{-5} \ M_\odot {\rm yr}^{-1}$, alternate with 
quiescent intervals where $\dot{M} \approx 10^{-8} \ M_\odot {\rm yr}^{-1}$. 
Both the duration of the outbursts, and the recurrence time, vary with 
the rate at which gas is supplied to the outer disc. However, to order 
of magnitude, for $\dot{M}_{\rm infall} \ale 10^{-6} \ M_\odot {\rm yr}^{-1}$ 
we find $t_{\rm outburst} \sim 10^4 \ {\rm yr}$, and 
$t_{\rm recur} \sim 10^5 \ {\rm yr}$.

\begin{figure}
\psfig{figure=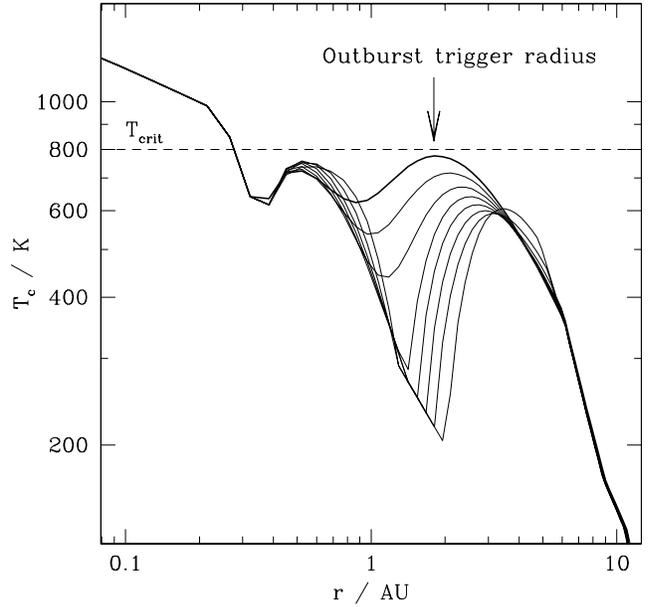,width=3.5truein,height=3.5truein}
\caption{Evolution of the disc central temperature during 
	quiescence. An outburst is triggered when the 
	central temperature in the layered region first 
	exceeds $T_{\rm crit}$ (dashed line), the temperature 
	at which the disc is well enough ionized to support MHD 
	turbulence. The curves show temperature profiles plotted 
	at intervals of $10^4$ yr, with the earliest timeslice 
	having the lowest temperature at 2 AU. The final timeslice,
	shown as the upper, bold curve, immediately precedes the 
	start of an outburst.}
\label{f3}
\end{figure}

The mechanism for these outbursts is shown in Figure {\ref{f3}}, 
which plots the central temperature during the quiescent interval 
leading up to an outburst. Following the end of one outburst, there 
is an extended layered region of the disc between $\approx 0.3$ AU 
and $\approx 3$ AU. This region is stable against gravitational 
instability, and supports only the small quiescent accretion rate.
Gas flowing inwards from the more active self-gravitating 
region of the disc thus accumulates in the quiescent region,
and the transition radius where the disc 
becomes self-gravitating moves inward. At the same time, the 
greater dissipation rate in the self-gravitating disc creates 
a local peak in the disc temperature, which also moves inwards.
A new outburst is triggered 
when this peak first exceeds $T_{\rm crit}$, thereby satisfying the 
condition for MHD turbulence to be restarted. Once the outburst begins 
thermal energy is rapidly advected inwards, triggering a 
transition of the entire inner disc into a hot turbulent state with 
high accretion rate. This continues until the reservoir of accumulated 
mass has been depleted, whereupon a cooling wave sweeps through the 
disc and returns it to the quiescent, layered, state.

The derived outburst durations are consistent with the expected 
viscous timescale in protoplanetary discs at radii of a few AU.
For an outburst triggered at a radius of approximately 2~AU,
the viscous timescale once the disc is in outburst, given by
\begin{equation} 
 t_\nu \simeq { r^2 \over \nu } = { 1 \over {\alpha \Omega} } 
 \left( {h \over r} \right)^{-2},
\label{eq_tnu1}
\end{equation}
where $h$ is the disc scale height, is approximately, 
\begin{equation} 
 t_\nu \approx 2 \times 10^4 
 \left( {\alpha \over 10^{-2}} \right)^{-1} 
 \left( {r \over {\rm 2 \ AU}} \right)^{3/2}
 \left( { {h/r} \over 0.05 } \right)^{-2} \ {\rm yr}.
\label{eq_tnu2}
\end{equation}  
This is only a rough estimate, but it is of the correct 
order of magnitude to match the numerical model.

The involvement of disc self-gravity means that relatively large
disc masses are required for the limit cycle to operate. In our models, 
$M_{\rm disc}$ fluctuates between 0.2 and 0.3 $M_\odot$. 
During an outburst, only a small fraction of the disc mass --- around 
20\% --- is accreted. In more realistic evolutionary
models, in which the mass infall rate declines with time, the disc will 
therefore be able to produce a handful of outbursts, with a few hundredths of 
a solar mass of gas accreted during each, before settling into a 
permanent quiescent state.

\subsection{Relation to FU Orionis events}

The most striking variability observed in pre-main-sequence 
stars occurs in FU Orionis events (e.g. Kenyon 1995; Hartmann \& 
Kenyon 1996), which are large 
amplitude outbursts in the system luminosity originating in the accretion disc. The 
statistics of these events are subject to considerable uncertainty, but 
the peak accretion rate during outbursts is of the order of 
$10^{-4} \ M_\odot {\rm yr}^{-1}$, while the duration is generally 
thought to be around $10^2$ yr. 

The most popular explanation for FU Orionis events is in terms 
of a disc thermal instability (e.g. Bell \& Lin 1994; Kley \& Lin 1999, 
and references therein), akin to that used to model dwarf novae (e.g. Cannizzo 1993). 
In the protostellar case, a thermal instability can only operate 
at extremely small disc radii, typically at less than 0.1~AU. Matching the 
observed timescales then
requires small values of $\alpha$ ($10^{-3}$ to $10^{-4}$), with
correspondingly large values for the disc surface density. As noted by Gammie (1999), 
it would be attractive to dispense with the need for a thermal 
instability by invoking an alternative limit cycle of a layered 
disc. This might be possible {\em if} the transition to the 
outburst state was triggered at the extreme inner edge of the layered 
region. Whether this happens depends on where in the quiescent 
layer mass preferentially accumulates, and the outcome is therefore 
sensitive to the detailed, and rather uncertain, physics of the layered disc.
In our model, in which matter is being added to 
the layered region primarily at large radius, the triggering occurs at the outer 
edge of the layered region, and the timescales are inconsistent with 
those of FU Orionis events. We note, however, that the accretion 
rates of $(1-10) \times 10^{-5} M_\odot {\rm yr}^{-1}$ that we 
obtain during our outbursts fall into the thermally unstable 
regime. The long 
outbursts obtained here could then feed shorter FU Orionis event 
of the sort calculated by Bell \& Lin (1994).

\subsection{Protostellar accretion history}

Continuous replenishment of the disc mass from infall is not a 
realistic model for protostellar accretion.
To explore how the accretion rate onto the star may evolve with time, 
we adopt a simple model for the infall of gas onto the disc. The 
details are model-dependent (see e.g. Larson 1969; Shu 
1977; Basu 1998), but at early times the 
infall rate is expected to be of the order of $\dot{M}_{\rm infall} \sim 
c_s^3 / G$, where $c_s$ is the sound speed in the collapsing 
cloud. For cloud temperatures $T \sim 10 \ {\rm K}$, this implies an 
infall rate of $\sim 10^{-5} \ M_\odot {\rm yr}^{-1}$. We take 
an initial infall rate of $2 \times 10^{-5} \ M_\odot {\rm yr}^{-1}$, 
and assume that this declines exponentially on the free fall timescale, 
$t_{\rm ff} = (3 \pi / 32 G \rho_{\rm cloud})^{1/2}$, where $\rho_{cloud}$ 
is the cloud density. We take $t_{\rm ff} = 10^5 \ {\rm yr}$. Other 
input parameters are as described earlier.

\begin{figure}
\psfig{figure=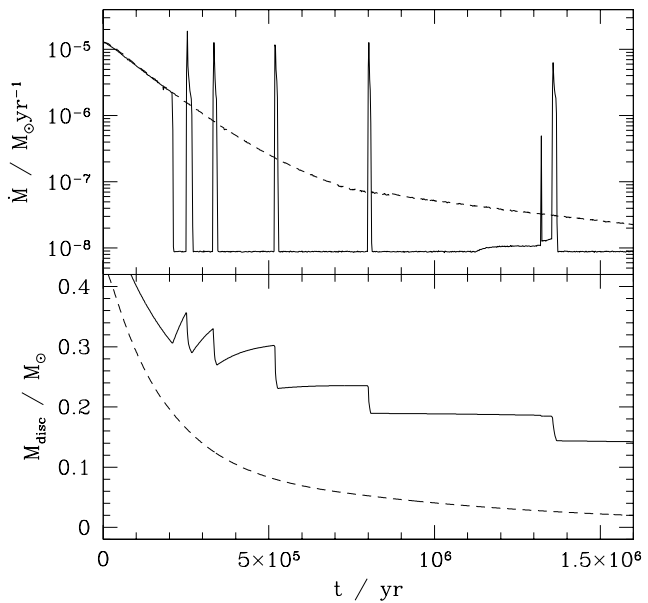,width=3.5truein,height=3.5truein}
\caption{Accretion rate and disc mass as a function of time. The solid 
	 curves are for layered disc models in which the mass infall 
	 rate onto the outer disc declines exponentially. For comparison with 
	 these, the dashed curves show the evolution of a model without 
	 a layered structure. This model has 
	 identical parameters, except that 
	 the disc is assumed to be viscous at all radii and central 
	 temperatures.}
\label{f4}
\end{figure}

With these parameters, Fig.~\ref{f4} shows the accretion rate for 
the layered disc model. For comparison, we also show the results 
from a viscous disc model which has identical parameters, but 
which has no layered region. This control model would be 
appropriate, for example, if angular momentum transport was 
driven by a purely hydrodynamic mechanism whose efficiency 
was independent of the disc temperature. We compute the 
control model by using the same code and infall history, 
but with $T_{\rm crit}$ set to zero.

Initially, when the 
infall rate is higher than around $2 \times 10^{-6} \ M_\odot {\rm yr}^{-1}$, 
both models support an accretion rate onto the star which tracks 
that with which matter is being added to the disc. Subsequently,
accretion through the layered disc occurs in outbursts with 
properties similar to those described earlier, but with 
increasing intervals between events. For this model, the last 
outburst before the disc became permanently quiescent 
occurred after 2 Myr. The accretion rate onto the star 
in the control model, on the other hand, declines smoothly 
with time. Similar results are obtained using other 
functional forms for the mass infall rate as a function 
of time.

The timescales of variability predicted by the model are 
much longer than any direct observational record. Indirect 
evidence for long timescale variability of young stars, however, is 
provided by studies of protostellar outflows, since the
rate of mass outflow via jets is widely believed to track 
the accretion rate. For example, observations by 
Reipurth, Bally \& Devine (1997) suggest that large amplitude 
variations in the outflow rate occur not only on the $10^2$ yr 
timescales characteristic of FU Orionis events, but also on 
much longer timescales of $10^4$ yr or greater. These longer 
timescales are characteristic of processes occurring at larger 
disc radii than thermal instabilities, and are consistent with 
instabilities, of the kind discussed here, originating in the 
layered disc region.

\begin{figure}
\psfig{figure=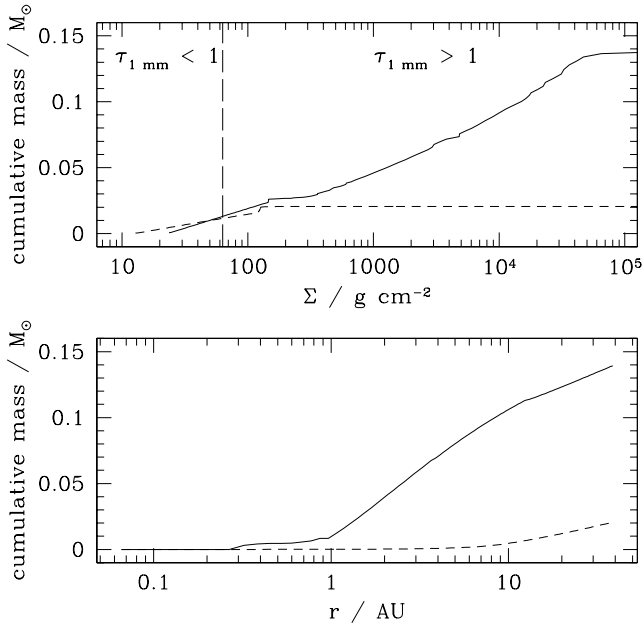,width=3.5truein,height=3.5truein}
\caption{Cumulative disc mass as a function of (upper panel) surface density $\Sigma$ 
	and (lower panel) radius. The solid curve shows 
	the layered disc model at $t=1.5 \times 10^6 \ {\rm yr}$. 
	For comparison, the dashed curve shows results for the 
	fully viscous model at the same time. The vertical dashed 
	line in the upper panel shows the approximate surface density below which the 
	disc would be optically thin at $\lambda = 1 \ {\rm mm}$. 
	Essentially all the additional mass present in the layered 
	disc is locked up in highly optically thick regions at 
	small radii. The amount of mass at $r > 10 \ {\rm AU}$ is 
	comparable in both the layered and the fully viscous model.}
\label{f5}
\end{figure}

The {\em outburst} accretion rate in the layered disc model is several 
orders of magnitude higher than that in the equivalent fully 
viscous disc model at $t \sim 10^6 \ {\rm yr}$. However, the 
time averaged accretion rate is substantially lower, so that at 
late times the disc mass in the layered model substantially 
exceeds that in the viscous model. We find 
that $M_{\rm disc} \age 0.1 \ M_\odot$ after 1 -- 2 Myr.
By this time the 
mass of the fully viscous disc has dropped to only 
$10^{-2} \ M_\odot$. As shown in Fig.~5, most of 
this extra mass is tied up in the quiescent layer at radii 
of the order of 1 AU, where the surface density 
exceeds the minimum mass solar nebula 
estimate (Hayashi, Nakazawa \& Nakagawa 1985) by about 
two orders of magnitude. 

\subsection{Compatibility with observations of protoplanetary discs}

A disc mass of a few tenths of a solar mass at early times 
is consistent with the upper end of the disc mass distribution 
inferred from mm-wavelength observations (Beckwith et al. 1990; 
Osterloh \& Beckwith 1995). However, the layered disc model predicts 
that a substantial mass -- at least a few hundredths of a solar mass -- remains 
in the disc throughout most of the typical Classical T Tauri 
disc lifetime (Strom 1995). Although recent measurements of 
the mass of ${\rm H}_2$ in debris discs indicate that several 
Jupiter masses of gas can persist beyond 
10 Myr (Thi et al. 2001), our predicted disc masses are 
in apparent contradiction with mm observations of 
many systems with disc masses estimated at $10^{-3} \ M_\odot$ or 
smaller. 

To check whether this is a serious problem for the layered disc model, 
we plot in Fig.~5 the cumulative distribution of disc mass as 
a function of surface density. Most of the `additional' mass 
in the layered disc, compared to the equivalent fully viscous 
control model, is locked up in the quiescent layer at small 
radii and high surface density. This material is highly optically 
thick, and thus would not contribute to the observed emission at 
mm wavelengths. For a range of dust models, 
the opacity of dust at $\lambda = 1 \ {\rm mm}$ is  
thought (Men\'shchikov \& Henning 1997; Beckwith et al. 1990) to be in the range 
$0.3 \ {\rm cm^2 g^{-1}} \ale \kappa \ale 10 \ {\rm cm^2 g^{-1}}$
(note that this is per gram of {\em dust}). For a gas to dust ratio of 100, 
this implies that an optical depth of unity at 1 mm is attained for 
$\Sigma \sim 10^2 \ {\rm g cm}^{-2}$. At this surface density 
threshold, the masses of optically thin gas in the layered and 
non-layered disc models are very similar. The masses 
of protoplanetary discs inferred from mm-wavelength measurements 
could thus substantially underestimate the true disc masses.

The accretion rates inferred for Classical T Tauri stars can 
also provide a test.
In the simplest version of the layered disc model,
the accretion rate during quiescence 
does not change over the lifetime of the protostellar phase. This rate 
is consistent with 
the order of magnitude inferred for accreting Classical T Tauri 
stars (Gullbring et al. 1998). There have also been 
suggestions that the observationally inferred accretion rate 
declines with increasing stellar age (Hartmann et al. 1998).
Although currently the scatter in measured 
accretion rates at a given assumed age is very large, and 
the age estimates themselves are subject to substantial 
uncertainties (Tout, Livio \& Bonnell 1999), this observation poses 
difficulties for the simplest layered disc model (Stepinski 1998).
We note, however, that changes in the quiescent accretion rate 
with time would be expected in more complete models, for example 
if the opacity from dust decreased with time due to grain 
growth. Including heating of the disc from the changing
stellar radiation would also lead to a decline in the 
quiescent accretion rate. 
 
\section{Planet formation}

The environment which a layered disc presents for planet formation 
differs in several respects from both previous viscous disc 
models (e.g. Lin \& Pringle 1990), and from static models such as the 
minimum mass solar nebula. At radii between a few tenths of 
an AU, and several AU, the gas near the disc midplane is 
almost always quiescent, unless the settling of cm sized 
particles itself drives additional instabilities (Goldreich \& Ward 1973; 
Cuzzi, Dobrovolskis \& Champney 1993). The low viscosity 
of the disc in this region would reduce the rate 
at which massive planets migrate inwards, and may allow 
planet-disc interactions to excite significant eccentricity 
(Papaloizou, Nelson \& Masset 2001).

Other differences include the  
disc at large radii remaining mildly unstable to 
gravitational instability 
for a relatively long time -- around a Myr -- because the reduced time-averaged accretion 
rate leads to a larger disc mass at late times when 
compared to viscous disc models. Outbursts, which 
persist for a substantial fraction of the disc lifetime, 
mean that much of the inner disc 
is expected to be subject to rapid heating and cooling 
episodes.
However, we find no evidence that these cooling waves drive the 
disc into a state where gravitational collapse ($Q \ale 1$) 
would occur.

The survival of solid bodies in protoplanetary discs is 
limited by the rate at which they migrate inwards relative 
to the gas. Migration can be rapid both for cm-sized bodies 
(e.g. Godon \& Livio 1999, and references therein), and is 
particularly problematic for low mass planets, for which migration 
occurs due to the influence of 
gravitational torques (Goldreich \& Tremaine 1979). In standard disc 
models, the migration timescale for Earth mass planets at a few AU 
can be as short as $10^4 - 10^5 \ {\rm yr}$, which leaves little 
time to assemble the cores of giant planets before the putative 
building blocks are consumed by the star. The rate of migration 
depends only very weakly upon the surface density profile
(Ward 1997), but {\em is} sensitive to the gradient of the 
central temperature, and can be halted if there exists a 
region of the disc where $T_c \propto r$. During the 
quiescent phase, the models we have discussed here possess 
such a non-monotonic central temperature profile at 
radii of 1-3 AU (Fig.~3), leading to 
the possibility of 
an enhanced probability of giant planet formation 
at those radii (Papaloizou \& Terquem 1999). Current radial 
velocity surveys are sensitive to massive planets at radii 
$r \ale 3 \ {\rm AU}$, i.e. within this region. If 
migration is as rapid as currently suspected, a 
consequence of the layered disc model would probably be fewer planets 
at larger radii than would be expected from an extrapolation 
from the current data.

\section{Summary}

In this paper, we have presented models for the evolution of magnetically 
layered protoplanetary discs. Given our present understanding of angular 
momentum transport in discs, this model represents the best 
guess for the structure of protoplanetary discs at radii 
of the order of 1 AU, where the gas is cool and poorly coupled 
to the magnetic field. The evolutionary models presented here 
suggest a number of important differences with non-layered viscous 
disc models.
\begin{enumerate}
\item[(i)]
The disc cannot be in a steady state for outer disc accretion rates 
in the range $10^{-8} \ M_\odot {\rm yr}^{-1} \ale \dot{M} \ale 
2 \times 10^{-6} \ M_\odot {\rm yr}^{-1}$. A limit cycle is 
obtained, in which heating when the layered region of the disc becomes self-gravitating 
periodically restarts MHD turbulence, leading to outbursts of 
accretion onto the star. All calculations to date of layered discs
obtain strongly episodic accretion, despite variations in the 
physical processes included in the models (Gammie 1999; Stepinski 1999). 
This is the most robust prediction of the model.
\item[(ii)] 
The duration of outbursts depends on where in the quiescent disc they 
are triggered. In our model, the outbursts are long, with 
duration $\sim 10^4 {\rm yr}$. They could drive repeated strong 
episodes of mass outflow from the inner disc, resulting in `pulsing' 
of the observed jets.
\item[(iii)]
The time-averaged accretion rate is reduced due to the bottleneck 
created by the layered region of the disc. As a result, the disc 
mass at late times is large, typically $\approx 0.1 M_\odot$ at 
1-2 Myr. Most of this mass is locked up in the quiescent layer 
of the disc at small radii.
\item[(iv)]
The central temperature is strongly modified, and often 
{\em increases} with radius, near the transition between 
the layered disc and the outer self-gravitating region.
This could slow the otherwise rapid rate of low mass planetary 
migration at radii $r \sim 1 - 2 \ {\rm AU}$. 
\item[(v)]
A low viscosity in the layered region of the disc affects 
the migration rate and eccentricity evolution of massive 
planets at the radii currently probed by radial velocity 
surveys.
\end{enumerate}
Current observations provide only very limited constraints on 
the properties of protoplanetary discs at the radii of 
greatest interest for planet formation. As a result, 
there remains substantial uncertainty in our knowledge 
of how these discs evolve. Models, such as this 
one, that attempt to include more realistic disc physics, 
can lead to very different environments for planet formation 
and migration than the highly simplified models usually 
considered. The most obvious observational 
predictions of the model are that 
some very young stars should be accreting at the low `quiescent' 
rate of the order of $10^{-8} \ M_\odot {\rm yr}^{-1}$, and 
that high accretion rate outbursts should continue, 
albeit with lesser frequency, during a substantial fraction of 
the Classical T Tauri phase.

\section*{Acknowledgments}

PJA thanks the Institute of Astronomy for their hospitality 
during the course of part of this work, and Charles Gammie, John Papaloizou  
and Henk Spruit for helpful discussions. This work 
was partially supported by the Training and Mobility of Researchers 
(TMR) program of the European Commission. ML acknowledges 
support from NASA Grant NAG5-6857. JEP is grateful to STScI for continuing 
support under their Visitors Program.

\end{document}